\documentclass[12pt]{iopart}
\usepackage{iopams}  
\usepackage{amssymb}
\usepackage{color}
\usepackage{graphicx}
\usepackage{tabularx}
\usepackage{dcolumn} % Align table columns on decimal point
\usepackage{bm}      % bold math
\usepackage{array}   % table making
\usepackage{amsfonts}
\usepackage{ulem}
\newcommand{\sect}[1]{Sect.~\ref{#1}}
\newcommand{\Sect}[1]{Section~\ref{#1}}
\newcommand{\fig}[1]{Fig.~\ref{#1}}
\newcommand{\Fig}[1]{Figure~\ref{#1}}
\newcommand{\eq}[1]{Eq.~(\ref{#1})}

\newcommand{\eqs}[1]{Eqs.~(\ref{#1})}
\newcommand{\Eqs}[1]{Equations~(\ref{#1})}
\newcommand{\eqnref}[1]{(\ref{#1})}
\newcommand{\tab}[1]{Table~\ref{#1}}

\newcolumntype{.}[1]{D{.}{.}{#1}}

\begin{document}
\title[vdW-DF study of electric field noise heating in ion traps caused by adsorbates]{vdW-corrected density functional study of electric field noise heating in ion traps caused by electrode surface adsorbates}

\author{Keith G. Ray$^1$, Brenda M. Rubenstein$^{1,2}$\footnote{K. G. Ray and B. M. Rubenstein contributed equally}, Wenze Gu$^2$ and Vincenzo Lordi$^1$}

\address{$^1$ Quantum Simulations Group, Lawrence Livermore National Laboratory, Livermore, CA 94550, USA}
\address{$^2$ Department of Chemistry, Brown University, Providence, RI, 02912, USA}
\ead{ray30@llnl.gov}

%  LLNL-JRNL-734203
%\date{\today}

\begin{abstract}
In order to realize the full potential of ion trap quantum computers, an improved understanding is required of the motional heating that trapped ions experience.  Experimental studies of the temperature-, frequency-, and ion--electrode distance-dependence of the electric field noise responsible for motional heating, as well as the noise before and after ion bombardment cleaning of trap electrodes, suggest that fluctuations of adsorbate dipoles are a likely source of so-called ``anomalous heating,'' or motional heating of the trapped ions at a rate much higher than the Johnson noise limit. Previous computational studies have investigated how the fluctuation of model adsorbate dipoles affects anomalous heating. However, the way in which specific adsorbates affect the electric field noise has not yet been examined, and an electric dipole model employed in previous studies is only accurate for a small subset of possible adsorbates. Here, we analyze the behavior of both in-plane and out-of-plane vibrational modes of fifteen adsorbate--electrode combinations within the independent fluctuating dipole model, utilizing accurate first principles computational methods to determine the surface-induced dipole moments.  We find the chemical specificity of the adsorbate can change the electric field noise by seven orders of magnitude and specifically that soft in-plane modes of weakly-adsorbed hydrocarbons produce the greatest noise and ion heating. We discuss the dynamics captured by the fluctuating dipole model, namely the adsorbate dependent turn-on temperature and electric field noise magnitude, and also discuss the model's failure to reproduce the measured 1/$\omega$ noise frequency scaling with a single adsorbate species. We suggest future research directions for improved, quantitatively predictive models based on extensions of the present framework to multiple interacting adsorbates.
\end{abstract}
\noindent{\it trapped ions, anomalous heating, surface adsorbates, decoherence, patch potential model\/}

\maketitle

% --------------------------------------------------------------------------------------
\section{Introduction}
\label{Introduction}

Due to their relative isolation, trapped ions allow unique access to the quantum world.  Their properties have long been exploited in atomic clocks and metrology.\cite{diddams2001optical,rosenband2008frequency,wineland2011quantum}  Looking forward, they hold the promise of becoming the principle components of the quantum computers of the future.\cite{kielpinski2002architecture,seidelin2006microfabricated}  Whether trapped ions will fulfill this role will ultimately depend upon their ability to remain coherent long enough to implement error correcting codes.\cite{bennett2000quantum,chiaverini2004realization} A significant contributor to the decoherence of trapped ions is motional heating, which causes the ions to be excited from their ground states into higher-lying vibrational states.\cite{wineland2011quantum,monroe2013scaling,hite2013surface} For instance, for a particular Molmer-Sorensen gate\cite{kirchmair2009deterministic} to achieve a fidelity of 99.9\%, the heating rate must be below $\sim$100~quanta per second.\cite{brownnutt2015ion} However, the heating rate in typical ion traps can be higher than 7000~quanta/s, depending on the particular trap configuration and operating conditions.\cite{hite2012100,hite2013surface} Because the magnitude of this noise greatly exceeds that attributable to the ubiquitous Johnson-Nyquist noise,\cite{brownnutt2015ion} this heating has been dubbed anomalous heating. In order to realize practical quantum computers, ion traps will need to be reduced in size.\cite{monroe2013scaling} 
Because heating rates scale as an inverse power of the ion--electrode distance, the ions in future architectures will be exposed to even greater noise as their distances from the electrodes are reduced. Developing strategies to minimize this noise will therefore be essential to the development of ion trap quantum computers.

Anomalous heating has been the subject of experimental investigation for several years.  In 2000, Turchette \textit{et al}.\cite{turchette2000heating} measured the heating 
rates of trapped ions in a variety of traps of different geometries and sizes, determining that the electric field noise scales with trap frequency as $\omega^{-\alpha}$, with $0.7 \leq \alpha \leq 1.5$, and ion--electrode distance 
as $d^{-3.8}$. Deslauriers \textit{et al}. reported similar results in 2006, measuring a $d^{-3.5}$ distance scaling of the noise,\cite{deslauriers2006scaling} while Boldin \textit{et al.} measured $d^{-3.8}$\cite{boldin2018measuring} and Sedlacek \textit{et al.} determined the distance scaling to be $d^{-4.0}$ or $d^{-3.9}$, depending on the temperature and assuming that the electric field noise is constant over the chip.\cite{sedlacek2018distance} We note that other trap geometries can yield a different distance-scaling, for example, Hite \textit{et al.} found $d^{-3.1}$ for a unique trap that positions the ion in close proximity to a sample material surface.\cite{hite2017measurements} The experimentally-determined distance scalings for planar traps are roughly consistent with those predicted by patch potential models of local regions of fluctuating potential,\cite{dubessy2009electric,low2011finite} which predict a $d^{-4}$ scaling for a planar trap with $d \gg \xi$, where $\xi$ is the fluctuation correlation length.  Such distance scalings are inconsistent with thermal electronic (Johnson) noise produced by metallic electrodes. 
Cryogenic temperatures have been shown to dramatically reduce the electric field noise, which ``turns on" between 40 and 70 K,\cite{labaziewicz2008suppression, chiaverini2014insensitivity} suggesting that the fluctuations are thermally activated. Furthermore, the noise is relatively insensitive to substrate 
choice.\cite{chiaverini2014insensitivity}  Laser cleaning has been demonstrated to reduce the electric field noise in one experiment,\cite{allcock2011reduction}
while in another experiment, without cleaning, the field noise increased over time,\cite{daniilidis2011fabrication} suggesting contaminants are a likely noise source. 
Using Auger spectroscopy, Hite \textit{et al.}\cite{hite2012100} found that carbon was present on their trap surfaces before cleaning and that Ar ion beam bombardment of Au electrodes can reduce the anomalous heating rate from 7000 to 43 quanta/s. As a result, they postulated that hydrocarbons may be a significant noise source.  Daniilidis \textit{et al.}\cite{daniilidis2014surface} also presented a study of the effect of ion bombardment, in this case
on Al/Cu electrodes, which showed that cleaning reduces carbon and oxygen Auger signatures as well as the heating rate in one device, from 200 to 3.8~quanta/s. They furthermore found that, 40 days after cleaning, portions of the oxygen and carbon were re-adsorbed, while the heating rate remained relatively small, demonstrating that the surfaces may not need to be atomically clean for the heating rates to be reduced and that the molecular identity of the adsorbates may be important.

Theoretical and computational investigations into microscopic surface sources of electric field noise include analytical models of randomly distributed fluctuating 
surface potentials,\cite{dubessy2009electric,low2011finite} simulations of adsorbate diffusion,\cite{PhysRevA.95.033407,JOOYA2018232} as well as first principles calculations of hydrogen on Au contacts\cite{safavi2011microscopic} and hydrogen 
on monolayers of He or N on Au.\cite{safavi2013influence}  Using a surface physisorption model and density functional theory, Safavi-Naini \textit{et al.}\cite{safavi2011microscopic} studied the heating rates produced by electrodes covered with hydrogen atoms. The hydrogen dipoles were modeled assuming that the atoms interacted with their image charges in the metal.\cite{antoniewicz1974surface} Based upon these assumptions, it was shown that a $\omega^{-1}$ frequency dependence arises in the dipole fluctuation spectra in certain frequency regimes.\cite{safavi2011microscopic, safavi2013influence}  This study demonstrated the promise of the patch potential model, however, the specific calculations used for surface--adsorbate interactions do not capture van der Waals forces and the dipole moment model does not capture the charge transfer arising from covalent and ionic bonding. Therefore, these theoretical treatments qualitatively fail for adsorbates for which these interactions play a role.
  In addition, the $\omega^{-1}$ frequency dependence obtained in that work was based upon model interaction potentials, electric dipole moments, and masses that do not correspond to many realistic adsorbates. 
 
In the following, we present a first principles methodology to obtain species-specific noise spectra, within the independent fluctuating dipole model, that is applicable to all adsorbates  and utilizes highly accurate computed adsorption potentials and dipole moments. We examine H, Ne, C, S, O, N$_2$, O$_2$, water (H$_2$O), carbon monoxide (CO), carbon dioxide (CO$_2$), methane (CH$_4$), acetylene (C$_2$H$_2$), ethene (C$_2$H$_4$), ethane (C$_2$H$_6$), and benzene (C$_6$H$_6$) adsorbates on Au using a van der Waals corrected density functional, the vdW-DF-cx.\cite{PhysRevB.89.035412}   We also consider H, S, CO, and CH$_4$ on Ag and S, CO, and CH$_4$ on Nb to study the influence of a different electrode material. 
After selecting for experimental relevance, several other species were examined primarily to demonstrate particular binding regimes, e.g., Ne as a vdW bound atom, or to establish trends, e.g., light hydrocarbons vs. heavier hydrocarbons. The resulting set of adsorbates spans a range of covalently bonded, weakly bonded, monoatomic, and molecular species with varying symmetries and permanent dipole moment{s}. 
To compute the dipole moment of the adsorbed species, we perform an integration of the 
DFT-calculated charge density, which accurately accounts for the effects of binding, the electrode surface dipole's interaction with the adsorbate, and the adsorbate's 
interaction with its image charge.   Notably, the DFT-calculated dipoles are qualitatively different than those calculated based upon an
image-charge-interaction-only model;\cite{antoniewicz1974surface} for example, we find that the DFT-calculated dipoles can change direction as a function of the adsorbate--electrode distance, which is not captured by the image-charge-interaction model. These more accurate interaction potentials and dipoles are then used in temperature- and 
frequency-dependent noise master equation calculations that yield dipole- and electric-field fluctuation spectra. For each adsorbed species, we use these spectra to calculate the resulting trapped ion heating rate as a function of temperature, trap frequency, adsorbate surface density, and ion--electrode distance. 

The results of these calculations reveal the successes and limitations of the independent fluctuating dipole model applied to realistic candidate adsorbates. The calculated electric field noise spectra for individual adsorbates exhibit white noise at low frequencies and cross over to a $1/\omega^2$ scaling at higher frequencies, contrary to the roughly $1/\omega$ spectrum observed experimentally near the trap frequency.  
For realistic adsorbates, a model beyond the independent dipole model, including, for example, intramolecular interactions as adsorbate densities increase or the dynamics of multiple adsorbed species, is required to capture the observed frequency dependence. 
 We show that the noise turn-on temperatures of hydrocarbon adsorbates are 30-50 K, close to the experimentally observed range of 40-70~K, while strongly bound individual atoms have turn-on temperatures above 200~K. The qualitative ranking of the possible effects on heating from different adsorbates is predicted, albeit as a lower bound, since only sub-monolayer coverages can be treated by this model and the effects of surface defects and other sources of heating are neglected.

We find that hydrocarbons produce the highest trapped ion heating rates, due to their weak surface interaction potentials. 
We also find that lower-energy in-plane vibrational modes can produce significantly greater electric field noise than out-of-plane modes for many species, and the complexity of the in-plane mode structure results in a $1/\omega^\alpha$ scaling, with $0 < \alpha < 2$, over a wider range of frequencies.
The analysis of the effects of in-plane modes suggests possible processes that produce the measured $1/\omega$ spectrum and future models to capture such dynamics.

% --------------------------------------------------------------------------------------
\section{Methods}

\label{Methods}

\begin{figure}
\centering
\includegraphics[scale=0.17]{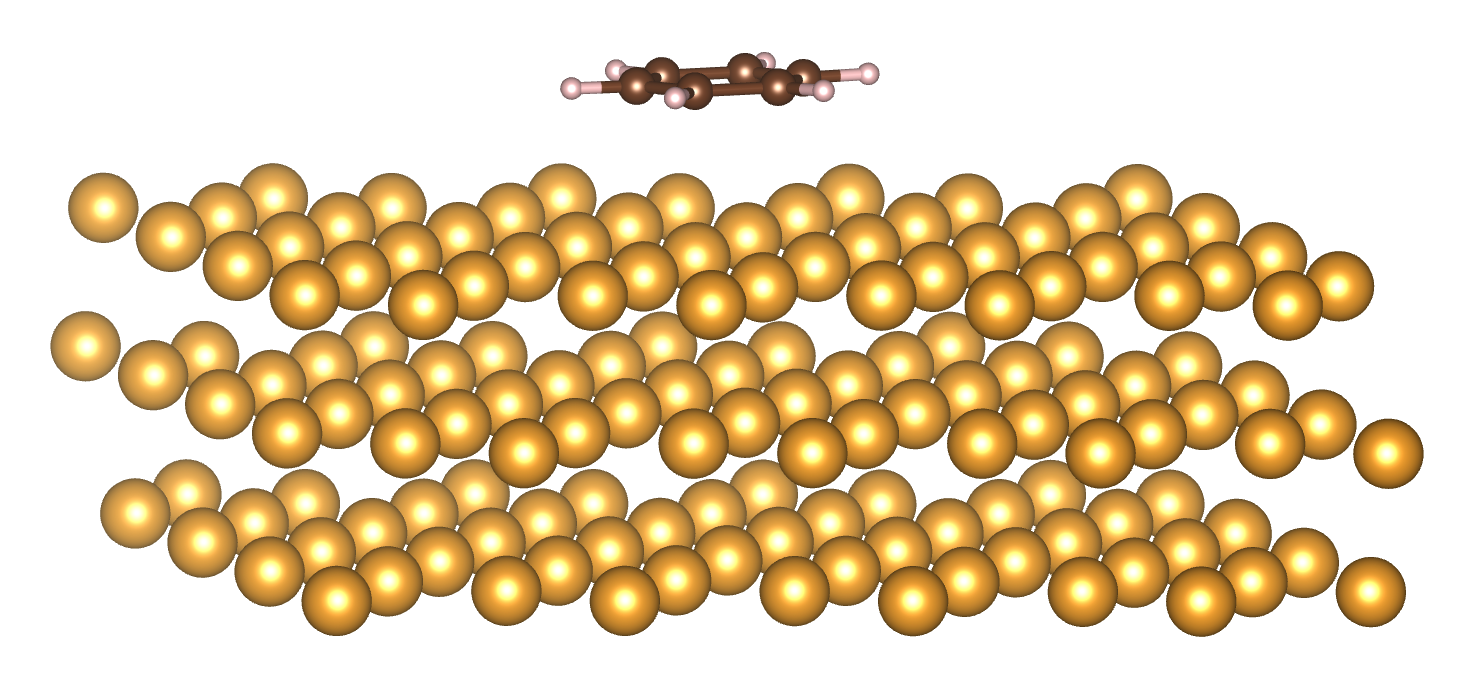}
\caption{Benzene adsorbed on a 144-atom, three-layer gold (111) slab. Au atoms are gold, C atoms are brown, and H atoms are beige.}
\label{benzene}
\end{figure}

To explore how specific adsorbates influence electric field noise spectra in the independent fluctuating dipole regime, we develop a multi-step approach based solely upon system parameters and \textit{ab initio} electrode--adsorbate interaction potentials to calculate noise spectra over a wide range of trap frequencies and temperatures.

Previous work\cite{safavi2011microscopic,turchette2000heating} has shown that when fluctuating electric fields couple to 
the motion of a trapped ion, the ion's heating rate, $\dot{\hat{n}}$, may be written as
\begin{equation}
\dot{\hat{n}} = \frac{q^{2}}{2 m_{I} \hbar \omega_{t}} S_{E}(\omega_{t}), 
\label{eqnn}
\end{equation}
where $\omega_{t}$ is the frequency at which the ion is trapped, $q$ is the ion's charge, $m_{I}$ is the ion's mass, and
$S_{E}(\omega) = \int_{-\infty}^{\infty} d{\tau} \langle \delta E(\tau) \delta E(0) \rangle e^{i \omega \tau}$ 
is the frequency spectrum of the fluctuating electric fields. 
While the source of these fluctuating electric fields has long been debated,\cite{brownnutt2015ion} they have previously been modeled by a family of patch potentials.\cite{dubessy2009electric,brownnutt2015ion,rossi1992observations,PhysRevLett.90.160403}   
These models posit that anomalous heating may be attributed to local variations of the potential above electrode surfaces, due either to 
surface adsorbates or different crystal orientations,\cite{rossi1992observations,PhysRevLett.90.160403} and can account for the 
experimentally-suggested $d^{-4}$ dependence of the heating rate with the distance $d$ between the electrode surface and the trapped ion.
In this work, we investigate the electric field noise spectrum, $S_{E}(\omega)$, arising from the dipole-dipole fluctuation spectrum, $S_{\mu}(\omega)$, of adsorbed atoms or molecules on the electrode surface.
The relation between these quantities is given by either
\begin{equation}
S_{E}(\omega) = \frac{3 \pi \sigma S_{\mu}(\omega)}{2 (4 \pi \epsilon_{0})^{2} d^{4}},
\label{eqnse}
\end{equation}
for out-of-plane dipole fluctuations, or 
\begin{equation}
S_{E}(\omega) = \frac{ \pi \sigma S_{\mu}(\omega)}{ (4 \pi \epsilon_{0})^{2} d^{4}},
\label{eqnse2}
\end{equation}
for in-plane dipole fluctuations, where 
\begin{equation}
S_{\mu}(\omega) = \int_{-\infty}^{\infty} d\tau \left( \langle  \mu_{z}(\tau)  \mu_{z}(0) \rangle -\langle \mu_{z}(0) \rangle^{2} \right) e^{i \omega \tau},
\end{equation}
$\mu_{z}(\tau)$ represents the magnitude of the dipole at time $\tau$, and $\sigma$ is the average areal density of 
adsorbates. Being able to accurately compute $S_{\mu}(\omega)$ is therefore fundamental to being able to accurately predict electric field noise spectra  and the associated ion heating rate, within this model.

In this work, we adopt an approach to compute the dipole-dipole spectrum similar in spirit to that employed by 
Safavi-Naini \textit{et al}.\cite{safavi2011microscopic,safavi2013influence} Here, it is extended using first principles methods to accurately capture the specific interactions between a number of experimentally relevant and theoretically illustrative adsorbates and the trap electrode surfaces.  We perform density functional theory (DFT) calculations to obtain the 
electrode--adsorbate interaction potential energy, 
$U(z)$, and electric dipole moment, $\vec \mu(z)$, for a variety of electrode--adsorbate combinations
at a range of heights, $z$, above the surface, from which $S_\mu(\omega)$ can be derived.  We employ the Vienna Ab-Initio Software 
Package\cite{vasp1,vasp2,vasp3,vasp4} (VASP) with a van der Waals corrected density functional (vdW-DF-cx),\cite{Dion:2004fk, PhysRevB.89.035412}  a plane-wave basis cutoff of 600~eV, and pseudopotentials from the projector augmented
wave (PAW) set.\cite{paw2} Additional details on the DFT calculations are available in the Supplemental Materials.\cite{supp} The dipole moment $\vec \mu(z)$ for a specific molecule at distance $z$ above the surface is computed directly by integrating over the calculated charge density, via
\begin{equation}
\vec \mu(z) = \int d^3 \vec r \,  \rho(\vec r, z) \vec r,
\end{equation}
where $\rho(\vec r, z)$ represents the charge density of the adsorbate-electrode system at position  $\vec r$ when the adsorbate is located at a height $z$ above the surface.
A representative DFT model for the interaction of a benzene molecule with a gold surface is depicted in Figure 1.  

The DFT-derived potential energy landscapes, $U(z)$, are used to calculate the energies and wavefunctions of each adsorbate's  bound quantum vibrational states, by solving the Schr\"{o}dinger equation for each potential. The average dipole in each vibrational state, $| i \rangle$, is then given by
\begin{equation}
\mu_{z,i} = \langle i|\vec \mu(z)|i \rangle . 
\label{dipvib}
\end{equation}

The transition rates that ultimately dictate the time-dependence of the dipole fluctuations are calculated from Fermi's Golden Rule, as written here for the absorption of a phonon from the metal electrode by the adsorbate:
\begin{numparts}
\begin{eqnarray}
 \label{fgr_group}
                 \Gamma_{i \rightarrow f}    =  \frac{1}{2}\frac{\omega}{2\pi\hbar\rho}  \left(\frac{2}{\nu^3_T}+\frac{1}{\nu^3_L}\right) \left| \mathcal{M} \right|^2 \left(n_s(\hbar\omega)+1\right),
        \label{fgr}\\
   \textrm{with} \ \mathcal{M} = \left\langle n_1^f, n_2, n_3 \right|    \frac{d}{dr_{a1}}U(\mathbf{r}_a,\{\mathbf{r}_i\})  \Big| n_1^i, n_2, n_3 \Big\rangle,
\end{eqnarray}
\end{numparts} 
 where
 $\hbar\omega$ is the energy difference between the initial and final adsorbate vibrational states, $\rho$ is the density of the electrode material, $\nu_T$ and  $\nu_L$ are the transverse and longitudinal sound velocities of the electrode material, $| n_1, n_2, n_3 \rangle$ denotes the adsorbate vibrational state with mode numbers indicated for each of the three directions ($f$ and $i$ denote final and initial states),  $\frac{d}{dr_{a1}}U(\mathbf{r}_a,\{\mathbf{r}_i\})$ is the derivative in the first direction  of the binding potential with the adsorbate at position $\mathbf{r}_a$ and electrode atom positions $\{\mathbf{r}_i\}$, and $n_s(\hbar\omega) = (e^{\hbar\omega/k_B T} - 1)^{-1}$ is the electrode material phonon distribution function. 
\Eqs{fgr_group} follow Safavi-Naini \textit{et al}.,\cite{safavi2011microscopic} but have been rederived to include adsorbate vibrations in multiple directions, account for different transverse and longitudinal electrode phonon dispersions, and include a factor of $\frac{1}{2}$ that arises from integrating over the projection of the surface phonon eigenvectors onto the direction of adsorbate vibration. 
We solve a master equation with these transition rates using kinetic Monte Carlo to obtain the dipole fluctuation spectrum, S$_{\mu}(\omega)$,
from which we then obtain S$_E(\omega_t)$ and the heating rate $\dot{n}$, via \eqs{eqnn}--\eqnref{eqnse2}.
Further technical details of our approach are given in the Supplemental Materials.\cite{supp}

% --------------------------------------------------------------------------------------
\section{Results and Discussion}
\label{Results}
\subsection{Adsorbate interaction potentials}
\label{int_section}
\begin{figure}
\centering
\includegraphics[scale=1.1]{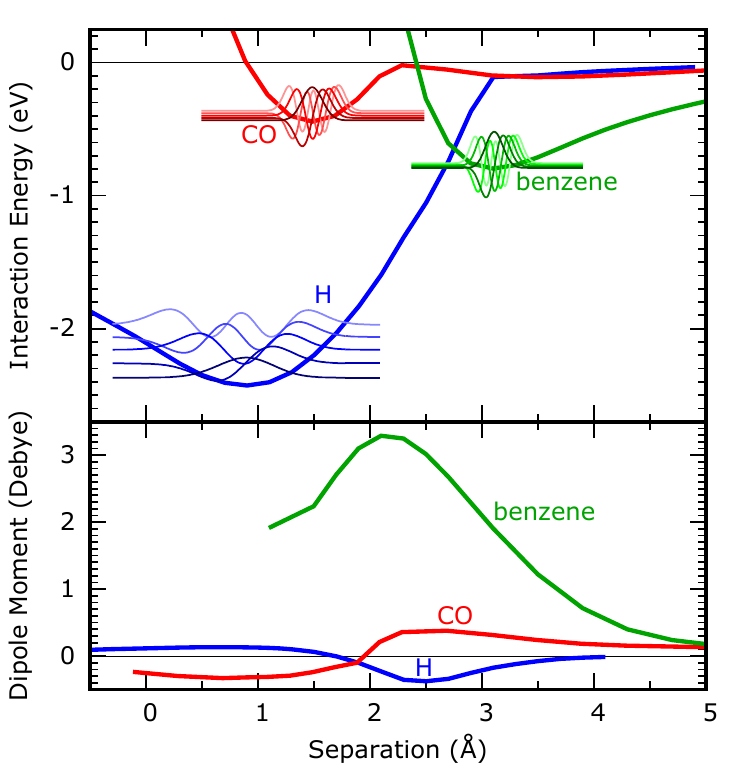}
\caption{Top: Interaction potentials and the first five vibrational wavefunctions for benzene, hydrogen, and carbon monoxide adsorbates on the Au(111) surface. The vertical position of each wavefunction indicates its energy. All energies are referenced to the infinite separation case at 0~eV. The wavefunctions are normalized  arbitrarily for maximum clarity. Bottom: Induced dipole moment as a function of adsorbate separation from the gold surface, where zero separation corresponds to the plane occupied by the surface layer of gold atoms. A positive dipole moment points away from the surface.
}
\label{FigureDipole}
\end{figure}

We aim to quantitatively calculate trapped ion heating rates due to fluctuating induced electric dipole moments created by adsorbed species on the electrodes. To achieve this, we require accurate adsorbate--electrode interaction potentials and induced dipole moments, which we calculate using the highly accurate 
vdW-DF-cx van der Waals corrected density functional. We consider H, Ne, C, S, O, N$_2$, O$_2$, water (H$_2$O), carbon monoxide (CO), carbon dioxide (CO$_2$), methane (CH$_4$), acetylene (C$_2$H$_2$), ethene (C$_2$H$_4$), ethane (C$_2$H$_6$), and benzene (C$_6$H$_6$) adsorbates on the Au(111) surface.  We also consider H, S, CO, and CH$_4$ on the Ag(111) surface and S, CO, and CH$_4$ on the Nb(110) surface, to compare the effect of the metal substrate.  The (111) surfaces are chosen for the fcc metals because they have the lowest surface energies while the (110) surfaces were chosen for the bcc Nb metal for the same reason.\cite{skriver1992surface}  The set of adsorbates considered include both weakly and strongly bound atomic species as well as a variety of hydrocarbons, representative of species likely occurring on trap surfaces.\cite{hite2012100,daniilidis2014surface}

For each adsorbate, we calculate the binding energy at several positions (and orientations for molecules), then compute the out-of-plane interaction potential for the lowest energy configuration.
For atomic adsorbates, the lowest energy site corresponds to a hollow area between three surface atoms.
For some adsorbates, we also computed the in-plane binding potential from the minimum energy path along the surface around the lowest energy configuration (see more details in \sect{transverse}).
The variety of binding distances and strengths are depicted in \fig{FigureDipole}, which shows the interaction potentials for H, CO, and benzene on Au.  The adsorbate--electrode interaction potentials for other combinations are presented in the Supplemental Materials.\cite{supp}  
\Fig{FigureDipole} also shows the first five adsorbate vibrational wavefunctions for H, CO, and benzene on Au, with the vertical position of each wavefunction indicating its energy.  

Among all the adsorbates we examine, the binding energies vary from $-0.031$~eV for He to $-4.96$~eV for C, while the equilibrium binding distances vary from 0.90~\AA \ for H to 3.10~\AA \ for benzene and 3.61~\AA \ for He.  (See the Supplemental Material\cite{supp} for the full set of values.)
Such binding distances and energies demonstrate the distinction between the chemically bound H and S species and the dispersion bound Ne, CH$_{4}$, and benzene adsorbates. In general, the adsorbate--substrate interaction may be a combination of dispersion, ionic, and covalent contributions. 
Accurate DFT treatment of this range of interactions requires the use of a vdW-corrected density functional. Previous treatments have neglected to employ such a high level of theory. As discussed in later sections, dispersion bound species produce the highest trapped ion heating rates and so this adsorption energy range is particularly relevant. These interactions
determine the full potential energy curve, which along with the adsorbate mass, define the adsorbate vibrational spectra. The energy splittings of the vibrational states dictate the temperature dependence of the noise spectra. For instance, heavier and more weakly bound adsorbates possess lower noise turn-on temperatures due to smaller vibrational level energy splittings.

\subsection{Surface-induced adsorbate dipole moments and comparison of models}

The form of the adsorbate dipole moment, as a function
of the distance from the electrode surface, has a significant impact on the electric field fluctuations at the trapped ion. Previous work\cite{safavi2011microscopic} had assumed that the adsorbate dipole
moment scales as $z^{-4}$ with distance, $z$, from the surface. This functional form was based on a 
perturbation theory analysis that is best suited for describing the interactions between light noble gases and metal surfaces.\cite{antoniewicz1974surface} 
However, many of the adsorbates of interest bind more strongly to the surface than helium or neon, so we applied a more general direct computation of the dipole moments using DFT.
Our approach allowed us to determine the applicability of this functional form and how this assumption affects the different adsorbates' resulting electric field noise spectra. 

\Fig{FigureDipole} shows for the illustrative cases of H, CO, and benzene that the $z^{-4}$ dependence does not hold for the dipole moments of species that strongly interact with the electrode (H) or for molecules with a permanent dipole (CO). There are substantial qualitative differences in the magnitude and, in some cases, the sign of the induced dipole.  For example, our DFT calculations show that the dipole moment of H on Au is positive (away from the surface) at small distances and switches to negative (toward the surface) at intermediate distances, 
before decaying to zero at large distances. 
This behavior arises, because close to the surface, electron charge density from the hydrogen is drawn toward the metal due to covalent chemical bonding and that charge density interacts with its image charge in the metal. Conversely, at larger distances where those effects are weak, adsorbate electrons are repelled from the surface due to the intrinsic surface dipole of the metal--caused by charge from under-coordinated metal surface atoms---extending into the vacuum.

\Fig{FigureDipole} further shows that the effect of the Au substrate on the (permanent) dipole moment of CO is opposite to that of H.  With CO, the dipole is negative at short distances and positive at longer distances, due to electron density being displaced toward the vacuum as the CO molecule moves toward the gold surface.  At larger separations for the CO--Au system, electrons are drawn toward the surface through either a charge transfer or an image charge interaction, producing a positive dipole.  Because isolated CO has a permanent dipole moment, the total moment (permanent + induced) does not decay to zero at infinite separation from the surface; for large separations, the dipole moment points away from the surface with its permanent value since the thermodynamically preferred binding orientation has the O atom further from the surface than C.
For benzene at distances $\gtrsim$2.5~\AA, the dipole moment follows a functional form similar to $z^{-4}$, as proposed in the literature.\cite{antoniewicz1974surface}  Without any dangling bonds, the image charge interaction of benzene with gold, which is captured by the model derived with perturbation theory,\cite{antoniewicz1974surface} dominates over charge transfer or bond formation. Generally, the induced dipole moment will depend on covalent bond formation, the magnitude and sign of charge transfer between the adsorbate and surface, and an image-charge interaction, in that order.

While the magnitude of the induced dipole moment of the adsorbate and electrode is significantly affected by the details of the adsorbate--electrode interaction, 
the ultimate effects on the noise spectra are determined by the dipole-dipole
correlation functions that involve the products of dipole moments (see the text after \eq{eqnse} and the results in the following sections).   
In fact, the sign change observed in the dipole moment versus distance for some adsorbates, \textit{e.g.,} CO, cancels in the 
final expressions for the noise spectra. 
On the other hand, the slope of dipole moment versus distance over the spatial range of the wavefunctions sensitively affects the noise, since fluctuations in the electric field arise from changes in the average dipole moment between different vibrational states of the adsorbate.
Thus, the difference of average dipole moments in different vibrational states, determined by integrating the distance-dependent dipole moment over the vibrational wavefunction via {\eq{dipvib}}, strongly affects the magnitude of the noise and can be used to assess the quality of the different dipole moment models.
\tab{Table6b} lists representative differences in the average dipole moment between vibrational states for H and CH$_4$ on gold, comparing the results of the explicit DFT-based dipole moment calculations with the approximate image-charge model ($\propto z^{-4}$). 
The more accurate DFT-derived dipole moments yield much larger differences for CH$_4$, by about a factor of 4, compared to the approximate model, suggesting that the corresponding prediction of heating rate will be significantly higher than estimates based on a previously-published model.
In contrast, the DFT-calculated dipole moment differences for vibrational transitions of H are much smaller than those given by the approximate model, by as much as 1500 fold, and predict much smaller heating rates than previously estimated.\cite{safavi2011microscopic}

\begin{table}
\begin{tabular}{c.{0.5}.{0.5}.{0.5}}
  \multicolumn{1}{c}{Vibrational State} &  \multicolumn{1}{c}{Model}  & \multicolumn{1}{c}{DFT-based}  & \multicolumn{1}{c}{Hopping Rate}  \\
  \multicolumn{1}{c}{Transition ($i \rightarrow j$)} &  \multicolumn{1}{c}{$\Delta \mu_{z,i\rightarrow j} $ (D)}  & \multicolumn{1}{c}{$\Delta \mu_{z,i\rightarrow j}$ (D)} & \multicolumn{1}{c}{at 300~K (Hz)}  \\
 \hline                          
  Hydrogen\\
  \hline
  0 $\rightarrow$ 1  &  7.63 &-0.0145 & 1.1\times10^{14} \\
 1 $\rightarrow$ 2  &  18.0 & -0.0171& 2.1\times10^{14} \\
  2 $\rightarrow$ 3 &  29.5 & -0.0187 & 2.9\times10^{14}\\
  3 $\rightarrow$ 4  & 12.2 & -0.0197  & 3.6\times10^{14}\\
    \hline      
%    \multicolumn{1}{r}{Vibrational} &  \multicolumn{1}{c}{Model}  & \multicolumn{1}{c}{DFT-Based} & \multicolumn{1}{c}{Hopping}  \\
%  \multicolumn{1}{r}{State CH$_4$} &  \multicolumn{1}{c}{$\Delta \mu_{z,i\rightarrow j} $  (D)}  & \multicolumn{1}{c}{$\Delta \mu_{z,i\rightarrow j} $  (D)}  & \multicolumn{1}{c}{Rate} \\
Methane \\
      \hline      
    0 $\rightarrow$ 1   &  -0.00961 &-0.0386& 1.9\times10^{11} \\
 1 $\rightarrow$ 2 &  -0.00965 & -0.0334 &  3.5\times10^{11}\\
  2 $\rightarrow$ 3 &  -0.00707 & -0.0228 & 4.7\times10^{11}\\
  3 $\rightarrow$ 4  & -0.00673 & -0.0207  & 5.7\times10^{11}\\
\end{tabular}
\caption{The predicted average dipole moment differences in the out-of-plane $z$-direction, $\Delta \mu_{z,i\rightarrow j} $, for H (top) and CH$_4$ (bottom) on Au(111), for each of the first four vibrational state transitions are compared for the approximate model of Antoniewicz\cite{antoniewicz1974surface}, utilized in recent studies of electric field noise,\cite{safavi2011microscopic,safavi2013influence} and direct DFT-based charge density integration to obtain the dipole moment as a function of adsorbate distance from the surface.  We observe that the approximate model predicts dipole moment differences much larger than DFT for H and much smaller for methane. 
We also list the corresponding hopping rates at 300~K for each transition in the last column.}
\label{Table6b}
\end{table}

In addition, the frequency dependence of the noise will be largely  dependent on the hopping rates between different adsorbate vibrational modes. \tab{Table6b} also lists the computed hopping rates between the first few vibrational modes for H and CH$_4$ on Au. 
The hopping rate for hydrogen, $\sim$100~THz, is larger than its ground state vibrational frequency under the harmonic approximation, $1.5$~THz. Therefore, the out-of-plane hydrogen modes should be overdamped and the response given by a broad resonance. Although the model breaks down for hydrogen, we include results for this adsorbate---as an upper bound---to compare with previous work in the literature. For other adsorbates, the hopping rate is smaller than the vibrational frequency. 

Similar to the case for out-of-plane motion, adsorbate in-plane vibrations cause the induced adsorbate dipole moment to fluctuate. As the adsorbate is displaced along the surface, the charge distribution is distorted both along the surface and into and out of the plane, thus we must consider and calculate the dipole moment fluctuation components both in the plane and out of the plane of the surface. We use the same procedure as described in \sect{Methods} to compute the components. Those results and the associated noise spectra produced by in-plane fluctuations are discussed later in \sect{transverse}.

%---------------------------
\subsection{Electric field noise frequency dependence}
\label{freqsection}

\begin{figure}
\centering
\includegraphics[scale=0.35]{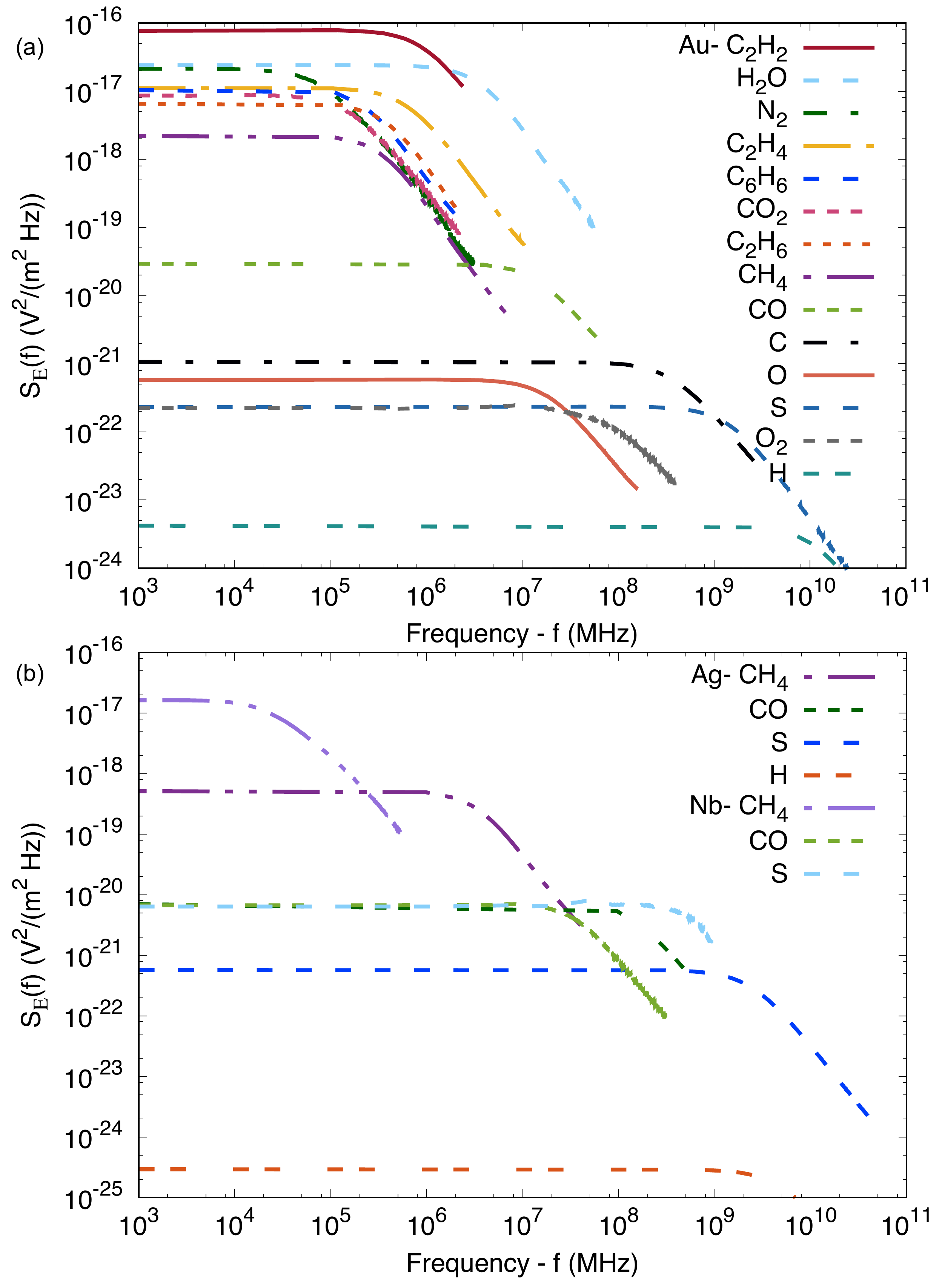}
\caption{Electric field noise spectra at 300 K for a selection of adsorbates on (a) Au(111) and (b) Ag(111) and Nb(110), at a coverage of $10^{18}$~m$^{-2}$ and a trapped ion at a distance of 40~$\mu$m. }
\label{spectracompare}
\end{figure}

Using the adsorbate--electrode interaction potential, the distance-dependence of the dipole moment, and the sound speeds in the electrode material, we calculate the wavefunctions, energies, and average dipole moments of each adsorbate vibrational eigenstate and obtain the electric field noise spectra associated with absorption and emission of phonons from the electrode as described in \sect{Methods}.
The results are summarized in \fig{spectracompare}.

\Fig{spectracompare} shows that  C$_2$H$_2$,  H$_2$O, C$_2$H$_4$, and C$_6$H$_6$ produce the largest magnitudes of electric field noise.
We may therefore generalize that hydrocarbons and water, representing weakly bound larger molecules, are more significant noise sources than the other adsorbates investigated here. In contrast, atomic adsorbates, including H and C \textit{in isolation}, do not produce significant noise in the model used in this work. Carbon monoxide molecules, bound more strongly than the hydrocarbons, but more weakly than the atomic adsorbates, produce an intermediate noise magnitude. Nitrogen molecules produce relatively high noise magnitudes, however, they are bonded very weakly to the gold electrode (120 meV) and are not likely to remain on the surface at room temperature.

Atomic chemisorbed species possess larger vibrational energy level separations and do not allow low energy transitions that are characteristic of the weakly bound adsorbates (see wavefunctions and caption in \fig{FigureDipole}).  These computational observations are consistent with measurements\cite{daniilidis2014surface} demonstrating that after electrodes are cleaned with an Ar$^+$ beam, Auger spectra taken of the samples show reduced carbon and oxygen features, and, after 40 days, these features partially return.  The return of these features, however, was found not to correspond to an increase in the noise. One explanation is that the form of the carbon is different: noise-causing hydrocarbons are present before cleaning and strongly bound individual atoms are present after cleaning.  Another experiment\cite{hite2012100} found that Ar$^+$-beam treatment of the electrodes reduced the trapped ion heating rate from 16,000 quanta/s to 134 quanta/s.  It then increased to 200 quanta/s over three days and remained constant (with 25$\%$) for 4 weeks in ultrahigh vacuum.

Comparing the effect of substrate metal, we find the electric field noise spectra produced by CH$_4$, CO, S, and H on the Ag(111) surface and CH$_4$, CO, and S on the Nb(110) surface are similar to those on Au(111). The relative magnitudes of the noise produced by each adsorbate is consistent between the three electrode metals and the noise power differs by roughly an order of magnitude at most, as seen in \fig{spectracompare}.

At the frequencies relevant for ion traps, 0.1--10~MHz, the electric field noise we calculate is virtually constant with frequency; as depicted in \fig{spectracompare}.  Beyond about 10$^4$--10$^{10}$~MHz, depending on the adsorbate, the noise decays as 1/$\omega^{2}$.    This behavior, which we call turnover, has been reported in the literature for another surface adsorbate model,\cite{safavi2011microscopic} where in that study, an intermediate S$_{E} \propto $ 1/$\omega$ frequency region was also found at particular temperatures and model adsorbate binding strengths.  A 1/$\omega^{\alpha}$  frequency dependence is observed in the experimental literature with exponents typically ranging from 0.7 to 1.5, although values as low as 0 and as high as 4 have been reported.\cite{brownnutt2015ion}  However, in our calculations of the out-of-plane adsorbate modes, the individual adsorbates exhibit 1/$\omega$ behavior only over a very small frequency range between the white noise and 1/$\omega^{2}$ noise frequency regions.    We note that we calculate wider frequency ranges corresponding to roughly 1/$\omega$ electric field noise scaling for in-plane vibrations, see \sect{transverse}.

The failure of the independent fluctuating dipole model applied to individual realistic adsorbates to capture the experimentally observed frequency scaling of the noise is a significant limitation. However, the dynamics from which 1/$\omega$ behavior arises experimentally may be represented by extensions to the current theory. Our model does not treat surface defects or multiple layers of adsorbates, complexity which would add variation in the binding energies and dipole moments, and produce a range of frequencies at which adsorbates transition from producing electric field noise with constant frequency dependence to 1/$\omega^{2}$. We were able to reproduce a 1/$\omega$ frequency scaling in model systems exhibiting several distinct transition rates not realized by the out-of-plane vibrational modes of the adsorbates we examine. These models suggest that a summation of the noise from a range of adsorbates with slightly different binding environments and therefore transition rates is likely to produce a spectrum with 1/$\omega$ scaling.

We further note that the calculation scheme used here is inappropriate when the energy of the adsorbate vibrational level transition exceeds that of the highest acoustic phonon in the electrode material ($\simeq$5~THz, or 21~meV, for Au). Above this energy, two-phonon processes must be considered, which have smaller transition frequencies and therefore may contribute less to the noise than first order processes. 
However, the species for which such second order transitions would be required include the strongly-bound adsorbates with stiff interaction potentials, such as the atomic adsorbates H, S, O, \textit{etc.}, which we have seen produce much less noise than adsorbates with more closely spaced vibrational transitions. 
 Therefore, the inclusion of second order transitions in addition to first order transitions does not change the qualitative results here,
particularly for the adsorbates with small vibrational energy splittings that produce the highest noise magnitudes, and comparisons among the considered adsorbates would not change.

\subsection{Comparative heating rates of selected adsorbates}
\label{heating}

\begin{figure}
\centering
\includegraphics[scale=0.355]{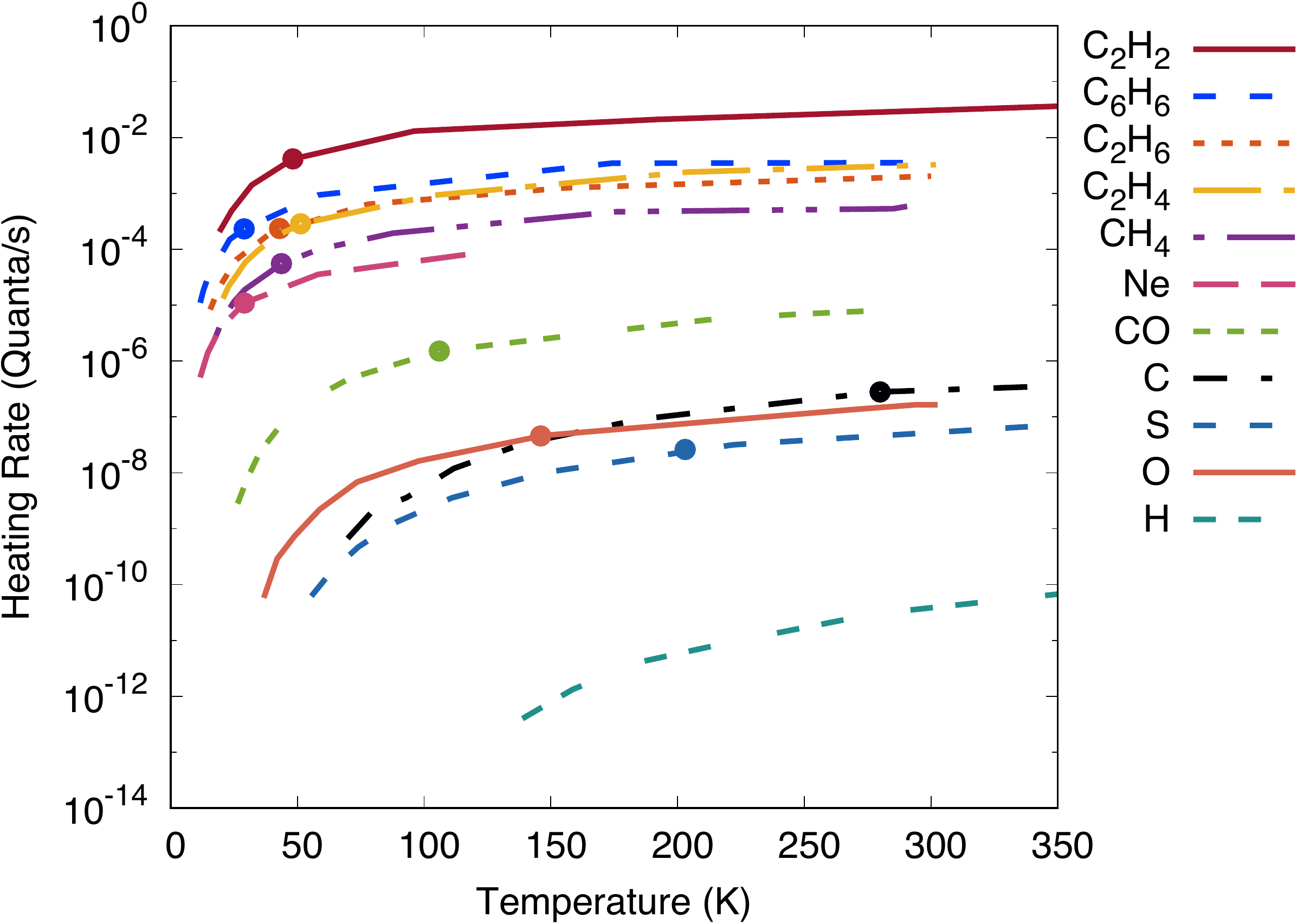}
\caption{Heating rates vs. temperature at 1~MHz for different electrode adsorbates on Au at a coverage of $10^{18}$~m$^{-2}$ and for a $^{40}\textrm{Ca}^+$ trapped ion at a distance of 40~$\mu$m.  Dots indicate the temperature corresponding to $\frac{1}{2}h \nu_{10}/k_{B}$ for each adsorbate; the value for H is 629~K, which is beyond the range of the plot.
}
\label{Tdepend}
\end{figure}

In \fig{Tdepend}, we convert the noise spectral data to a representative plot of the predicted trapped ion heating rate versus temperature, 
for a typical presumed trap consisting of a $^{40}\textrm{Ca}^+$ ion at a distance of 40 $\mu$m from a gold surface with adsorbate coverage of  $10^{18} $~m$^{-2}$ (approximately one adsorbate per square nanometer), and a trap frequency of 1~MHz. This frequency is well in the white noise region for all adsorbates studied and this still makes for a valid relative comparison of the heating strength of each adsorbate within the model employed. The chosen coverage corresponds to less than $< 0.4$ monolayer for each adsorbate, appropriate for the independent oscillator model used here, which neglects adsorbate--adsorbate interactions.  For reference, monolayer coverages for the species considered range from $\simeq 3\times10^{18}$~m$^{-2}$ to $>10^{19}$~m$^{-2}$, depending on the adsorbate.  The heating rate depends on the parameters mentioned in this paragraph: ion distance, mass, charge, and adsorbate coverage, as given in \eq{eqnn} and \eq{eqnse}.

At 300 K and with a $^{40}\textrm{Ca}^+$ ion distance of 40 $\mu$m, we find that C$_2$H$_2$ molecules produce heating rates $\sim$3$\times10^{-2}$ quanta/s for a surface density of 1~molecule/nm$^2$. This heating rate may be compared to the 3.8~quanta/s measured by Daniilidis, \textit{et al.}\cite{daniilidis2014surface} on a trap cleaned with an Ar$^+$-beam and consisting of a $^{40}\textrm{Ca}^+$ ion trapped 100~$\mu$m above the electrode ($\simeq$1 MHz trap frequency) and the 43~quanta/s measured by Hite, \textit{et al.}\cite{hite2012100} for a $^{9}\textrm{Be}^+$ ion trapped 40~$\mu$m above the electrodes (3.6~MHz trap frequency). 
Rescaling our results for the different trapped ion masses, heights, and frequencies used in the experiments, our calculated C$_2$H$_2$-produced ion heating rate is equivalent to 7.7$\times10^{-4}$~quanta/s for parameters corresponding to the Daniilidis, \textit{et al.} experiment and 1.3$\times10^{-1}$~quanta/s  for parameters corresponding to the Hite \textit{et al.} experiment.  Therefore, the predictions of heating rate within the independently fluctuating adsorbate model presented here, for out-of-plane vibrations at less than a monolayer of coverage, are considerably less  than those observed experimentally, and additional effects (some discussed below) are needed to fully explain the observed heating rates. In addition, we point out that the results presented here represent lower bounds for the predicted heating rates, since only a subset of possible mechanisms are included, and below we discuss several possibilities for expanded model complexity that can better correspond to experimental observables. 

In \sect{freqsection}, we showed that the out-of-plane adsorbate modes generate white noise up to $\sim$10$^4$--10$^{10}$~MHz, depending on the adsorbate, and that adsorbates exhibiting lower frequency modes tend to produce higher magnitude noise overall.  Therefore, if other lower frequency adsorbate modes are identified that continue this trend in the relationship between electric field noise magnitude and frequency, we expect to find higher heating rates at the relevant 0.1--10~MHz frequencies. Accordingly, the same future model extensions introduced in the previous section to capture the experimentally observed 1/$\omega$ frequency scaling---\textit{i.e.,} multiple layers of adsorbates, mixtures of different adsorbates, surface disorder, disorder in adsorbate binding orientations, and soft intramolecular vibrations---may also produce the experimentally observed heating rate magnitudes. In \sect{transverse}, we take a step in this direction by calculating the effects of softer in-plane vibration modes for several adsorbates, which produce higher heating rates and different frequency scaling than the out-of-plane vibrations.

%------------------
\subsection{Temperature dependence of electric field noise}
\label{sec_temp}

As discussed in \Sect{Introduction}, experiments on ion traps have shown that noise magnitude 
is strongly dependent on temperature,\cite{brownnutt2015ion} 
with indications that the noise may show an onset temperature in the tens of Kelvin,
above which there is significantly more noise than at lower temperatures.\cite{labaziewicz2008suppression, chiaverini2014insensitivity}

\begin{table}
\begin{tabular}{r.{2.5}.{2.5}.{2.5}.{2.5}}
\multicolumn{1}{c}{Adsorbate--Electrode} &  \multicolumn{1}{c}{$E_{1}-E_{0}$ (meV)}   &  \multicolumn{1}{c}{$ \frac{1}{2}h \nu_{10}/k_{B}$ (K)}  \\
\hline
Ne--Ag       &  4.41  &   25.7   \\
N$_2$--Au       &  4.93  &   28.8  \\
Benzene--Au  &  4.98  &   29.0     \\
Ne--Au       &  4.98 & 29.1\\
CH$_4$--Nb &  5.28     &    30.9    \\
CO$_2$--Au        &  5.38  &   31.4   \\    
CH$_4$--Au &  6.38     &    37.2    \\
C$_2$H$_6$--Au        &  7.38  &   43.1   \\
CH$_4$--Ag  &  7.49     &    43.7    \\
C$_2$H$_2$--Au        &  8.25  &   48.2   \\
C$_2$H$_4$--Au        &  8.79  &   51.3   \\
He--Au       &  9.93 & 58.0\\
H$_2$O--Au        &  14.5  &   84.8   \\    
CO--Ag        &  16.3  &   95.2   \\     
CO--Au        &  18.1  &   106   \\   
O$_2$--Au        &  23.6  &   138   \\     
O--Au        &  25.0  &   146    \\
CO--Nb        &  31.7  &   185   \\     
S--Ag        &  34.5  &   201    \\      
S--Au        &  34.7  &   203    \\      
C--Au        &  48.0  &   280    \\
S--Nb        &  81.2  &   474   \\     
H--Au        &  108  &   629   \\ 
H--Ag        &  122  &   715    \\ 
\end{tabular}
\caption{The energy difference between the ground and first excited vibrational states for  the out-of-plane motion of a variety of adsorbate--electrode combinations and the corresponding activation temperature at which higher vibrational states begin to be
occupied for each system. The combinations are ordered by increasing temperature. 
As illustrated in \fig{Tdepend}, higher vibrational states begin to be occupied at $T \simeq \frac{1}{2}h \nu_{10}/k_{B}$.}
\label{Table9}
\end{table}
One of our key findings is that the calculated dipole-dipole noise spectra at higher temperatures are several orders of magnitude larger than those obtained 
at lower temperatures. 
Overall, the substantial change in the spectra above $T =  \frac{1}{2}h \nu_{10}/k_{B}$ suggests that, much as in recent experiments,
our simulated spectra also manifest ``onset temperatures.'' 
Because the onset seen in our simulations occurs near $ T = \frac{1}{2}h \nu_{10}/k_{B}$,
different onset temperatures are expected for different adsorbate--electrode combinations, as listed in \tab{Table9} and indicated by the bold dots in \fig{Tdepend}. 
We note the onset temperature is directly proportional to the energy difference between the adsorbate's ground and first excited vibrational states, also listed in \tab{Table9}, which in turn depend on the adsorbate's mass and the depth of the adsorbate--electrode interaction potential. 

For example, for the CH$_4$--Au system, for $T \lesssim 37$~K $= \frac{1}{2}h \nu_{10}/k_{B}$, where $k_{B}$ denotes Boltzmann's constant,  
$h$ is Planck's constant, and $\nu_{10}$ denotes the frequency difference between the ground and first excited states, the system is 
roughly two-level in nature: only the lowest two vibrational states are occupied to any appreciable degree. On the other hand, 
for $T \gtrsim 37$~K, the system occupies a multitude of excited states; by 450~K, all bound vibrational states are equally occupied.

 As can be seen from 
\tab{Table9}, noble gas adsorbates exhibit the smallest energy difference between their first two bound vibrational states, and thus also  the smallest onset temperatures, largely owing 
to their weak  interactions with the electrode surface. In contrast, the S--Au, C--Au, and H--Au combinations have far stronger surface interactions leading to 
much larger energy differences and onset temperatures.

Interestingly, \fig{Tdepend} clearly shows the adsorbates associated with the highest ion heating rates (generally hydrocarbons and Ne)
exhibit onset temperatures all clustered around 30--50~K, within the range where several experiments have found that noise spectra change qualitatively ($\sim$40--70~K)\cite{labaziewicz2008suppression,chiaverini2014insensitivity}.
While Ne is not expected to be a common contaminant on electrodes, hydrocarbon fragments have long been discussed in the literature as 
possible sources of heating.\cite{hite2012100}

%----------
\subsection{In-plane vibrational modes}
\label{transverse}

\begin{figure}
\centering
\includegraphics[scale=0.345]{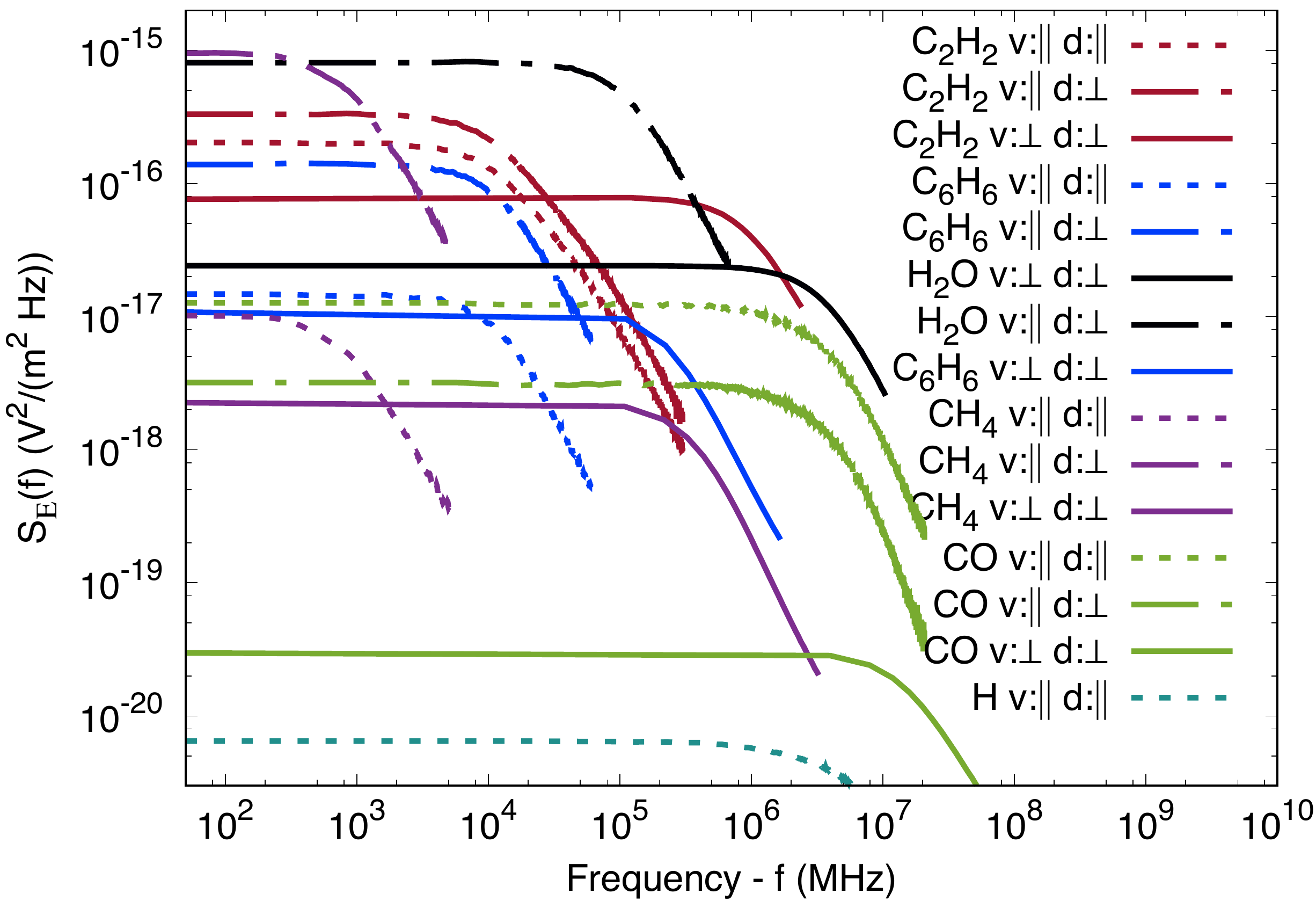}
\caption{Electric field noise spectra vs. trap frequency at 300 K for the in-plane vibrations, denoted v:$\parallel$, of a selection of electrode adsorbates at a coverage of $10^{18}$~m$^{-2}$ and a trapped ion at a distance of 40 $\mu$m.  Vibrations parallel to the surface produce adsorbate dipole moment fluctuations both in the in-plane and out-of-plane directions, denoted by d:$\parallel$ and d:$\perp$.  For comparison, the results for out-of-plane vibrations v:$\perp$, from \fig{spectracompare}(a), are also plotted for these adsorbates.
In general, ``v'' and ``d'' denote the direction of vibration and dipole moment component, respectively.
}
\label{spectracomparetransverse}
\end{figure}

Adsorbates on the electrode surface exhibit in-plane vibrational modes that also can produce dipole fluctuations that contribute to the electric field noise.  We calculate these modes for C$_2$H$_2$, C$_6$H$_6$, CH$_4$, H$_2$O, CO, and H on Au.  When an adsorbate is displaced in the plane, the induced dipole moment changes in both the in-plane and out-of-plane directions, which we denote by $\parallel$ and $\perp$, respectively.  
The electric field noise spectra resulting from in-plane vibrations are shown in \fig{spectracomparetransverse}, which also compares the spectra from out-of-plane vibrations.
For CH$_4$ on Au, we find that the electric field noise produced by the dipole fluctuations induced by this lateral motion is a factor of 415 greater than that produced by the out-of-plane motion, representing the highest magnitude of electric field noise generated by an adsorbate in this study. The equivalent heating rate from this lateral motion of CH$_4$ is 0.28~quanta/s at 300~K for a coverage of 1/nm$^2$ and the same assumed trap parameters as in \Sect{heating} and used in \fig{Tdepend}. 
As a frame of reference, this heating rate converts to 
1.2~quanta/s for a $^{9}$Be$^+$ trapped ion at 3.6 MHz and 40~$\mu$m above the surface, which is one order of magnitude smaller than the value measured experimentally with a cleaned $^{9}$Be$^+$ trap (43 quanta/s).\cite{hite2012100}  Thus, we observe that the dilute independent dipole model used here is consistent with predicting a lower bound for the heating, even in the case where the electric field noise is maximal based on the choice of adsorbate and vibrational modes.

\begin{figure}
\centering
\includegraphics[scale=0.345]{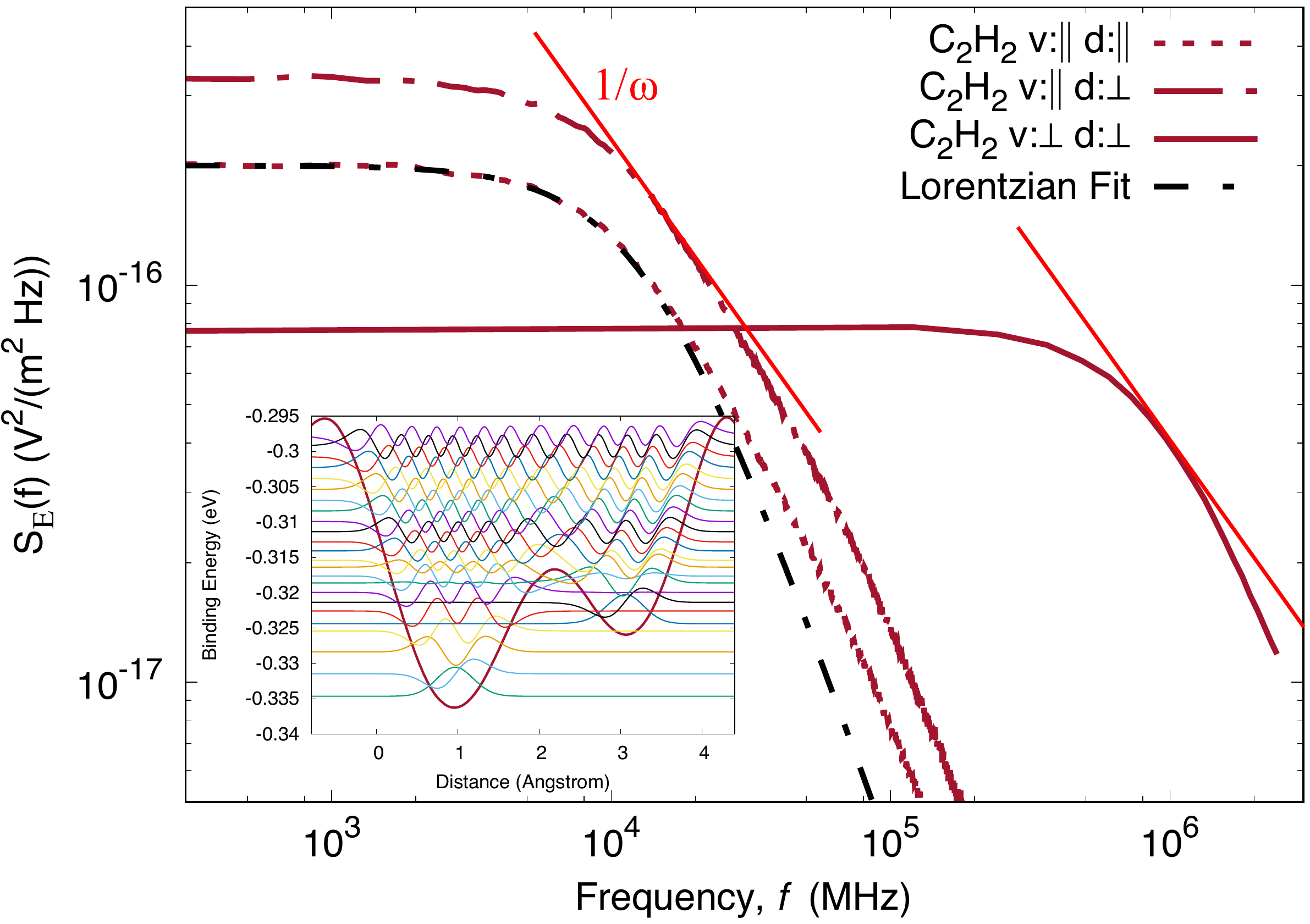}
\caption{Electric field noise spectra vs. trap frequency at 300 K for the in-plane vibrations, v:$\parallel$, and out-of-plane vibrations, v:$\perp$, of C$_2$H$_2$ at a coverage of $10^{18}$~m$^{-2}$ and a trapped ion at a distance of 40 $\mu$m.  Dipole moment fluctuations in the in-plane and out-of-plane directions are denoted by d:$\parallel$ and d:$\perp$.  Lorentzian and $1/\omega$ fits are included for comparison. Inset: the in-plane C$_2$H$_2$ potential, with two distinct wells, and several wavefunctions corresponding to the vibrational modes of C$_2$H$_2$ in this potential.  The vertical offsets of the wavefunctions denote their energy, on the same scale as the potential.
}
\label{C2H2}
\end{figure}

Generally, we find that the in-plane modes produce greater electric field noise than out-of-plane modes.
Compared to the out-of-plane modes for C$_2$H$_2$, C$_6$H$_6$,  H$_2$O, CO, and H, the lateral modes produce electric field noise that is 4.4, 12, 33, 420, and 402 times as great, respectively.  The greater electric field noise and heating rates are due to the soft in-plane vibrational modes created by the weak corrugation of the in-plane potential from the atomic structure of the surface, as illustrated for the case of C$_2$H$_2$ in \fig{C2H2}. The inset of \fig{C2H2} shows the in-plane potential corrugation and the associated vibrational wavefunctions. This potential energy landscape contains regions where the potential changes by less than 20~meV over a distance of greater than 2~\AA, which represents significantly smaller variations compared to the energy scale of the out-of-plane binding potential. 
For example, the potential energy of C$_2$H$_2$ changes by 100~meV when moved 1~\AA\ further from the surface than its equilibrium position.
 The softer in-plane modes have smaller energy splittings and higher occupations at the given temperature, compared to the out-of-plane modes, leading to the generally higher electric field noise.

\label{transverse}

In addition to the electric field noise magnitude, the frequency dependence  also manifests differently for the in-plane vibrational modes.
For the in-plane modes of C$_2$H$_2$, C$_6$H$_6$, CH$_4$, H$_2$O, CO, and H on Au, the transition from white noise to 1/$\omega^{2}$  occurs at much lower frequencies, between 10$^2$--10$^7$~MHz, compared with 10$^4$--10$^{10}$~MHz for the out-of-plane modes of the same adsorbates.  Furthermore, there exist frequency ranges over which the electric field noise scales as 1/$\omega^\alpha$, where $0 < \alpha < 2$.  For instance,  electric field noise scaling with approximately 1/$\omega$ is exhibited for the in-plane vibrational modes of C$_2$H$_2$ for frequencies between 1.0$\times$10$^{4}$ and 3.0$\times$10$^{4}$~MHz, as depicted in  \fig{C2H2}. In experiments, the electric field noise has been found to scale as roughly 1/$\omega$ over larger frequency ranges, however, the result we obtain for C$_2$H$_2$ is significant because this is the largest frequency range calculated in this study over which the noise scales like 1/$\omega$. In contrast, the out-of-plane modes for C$_2$H$_2$ produce electric field noise that more abruptly transitions from white noise to 1/$\omega^2$.  
We attribute this qualitative difference in the frequency scaling of the noise from in-plane modes to the two-well corrugation potential governing the in-plane motion, as shown in the inset of \fig{C2H2} along with the C$_2$H$_2$ vibrational wavefunctions.  Each well has a different curvature, producing a different energy level spacing; furthermore, at higher energies, the states occupy both wells and have a much smaller energy level spacing.  The resulting combination of three distinct energy level spacings is likely responsible for the more complex frequency dependence, with a distinct 1/$\omega$ region in between the flat and 1/$\omega^2$ regions, as suggested by model calculations described in \Sect{freqsection}. To demonstrate that this behavior is indicative of the action of multiple distinct hopping rates, a
 Lorentzian fit corresponding to the frequency response of noise from an adsorbate with a single hopping rate has been included for comparison in \fig{C2H2}, exhibiting a faster tail at high frequencies than seen in the calculated spectrum from in-plane vibrations.

The origin of the 1/$\omega$ noise scaling calculated in this model suggests that fluctuating dipole systems with more complicated and degenerate energy landscapes extend this scaling to lower frequencies and produce more electric field noise.   Higher coverage heterogeneous adsorbate systems on defective surfaces may extend this scaling to the lower frequencies observed experimentally.

% --------------------------------------------------------------------------------------
\section{Summary}
\label{Summary}

In this paper, we employed an independent fluctuating dipole model to explore the effects of a wide range of adsorbates on motional heating in ion
traps. In order to better understand the limitations and applicability of this model, we strove to make the microscopic
inputs into our implementation of the independent fluctuating dipole model as accurate as possible by relying on first principles DFT calculations of the relevant parameters for a variety of realistic adsorbates. We found that
using interaction potentials derived from vdW-corrected DFT calculations can dramatically alter the shape and magnitude of the noise spectra by substantially altering
the state-to-state transition rates that enter the electric dipole
autocorrelation functions. 
We  also found that the functional form commonly used in this context to estimate adsorbate dipole moments as a function of their distance above a
metallic surface is often not just quantitatively, but qualitatively incorrect. For strongly-bound adsorbates such as sulfur, oxygen, and carbon, 
the dipole moment calculated using DFT typically is positive close to the surface, then becomes negative further from the surface, and finally returns to zero at large separations---a behavior dramatically different from that
predicted by a previously-employed $z^{-4}$ ansatz.\cite{antoniewicz1974surface}  This discrepancy suggests that 
electrode--adsorbate interactions are more complex than previously thought and that special care is needed to 
accurately model dipole moments and their dynamics for studies of this nature. 

We find that the frequency dependent electric field noise spectrum produced by adsorbate out-of-plane vibrational modes exhibits white noise for low frequencies and a 1/$\omega^2$ scaling at high frequencies. The crossover between these behaviors for this independent fluctuating dipole model occurs between 10$^4$ and 10$^{10}$~MHz.  The intermediate frequency region is not wide enough to explain the 1/$\omega$ scaling around 1--10~MHz observed in experiments.  This significant discrepancy clearly signals that there should be dipole fluctuations at long timescales not captured by the independent dipole model, as currently presented in the literature. We hypothesize that a mixture of adsorbates and multiple adsorbate modes, each characterized by their own frequencies and energy scales, might exhibit an overall 1/$\omega$ electric field noise behavior.  Consistent with this hypothesis, we find that the frequency scaling of electric field noise from in-plane vibrational modes of some adsorbates follows a 1/$\omega$ trend over a slightly wider frequency range than the out-of-plane modes and that the transition from white noise to a 1/$\omega^\alpha$ scaling occurs at lower frequencies, between 10$^2$ and 10$^7$~MHz. The presence of multiple distinct transitions for the in-plane modes of some adsorbates is responsible for this behavior, and we hypothesize that interactions between multiple adsorbates could produce 1/$\omega$ noise frequency scaling at experimental frequencies. Collective motion of a high density of adsorbates may also result in lower frequency 1/$\omega$ behavior.

We demonstrate that our independent fluctuating dipole model is 
capable of manifesting behavior that is reminiscent of an onset temperature for anomalous trapped ion heating. When the ratio of $k_{B} T$ to the energy difference between the 
first two bound vibrational states of the interaction potential exceeds one half, the decay time of our simulated dipole-dipole correlation functions
rapidly decreases, leading to a significant deviation in the form of the spectra from that expected from a two-state model. Our model furthermore
predicts that the temperature at which the noise for hydrocarbons begins to manifest these changes is roughly in line with the onset temperatures
observed in experiments. While non-metals like S and O bind too strongly, hydrocarbons seem to possess the optimal combination of binding strength and mass  to reproduce experimental observations of the temperature dependence of noise.   In addition, at a relatively dilute coverage of one adsorbate per nm$^2$, appropriate for non-interacting adsorbates, the heating rate magnitude that we predict from the in-plane vibrational modes of CH$_4$ is within a factor of 35 of the lowest heating rates observed experimentally at 300 K for cleaned traps. Calculations that treat higher coverages and introduce adsorbate interactions may predict heating rates consistent with the experimentally measured rates.

Future work will investigate the use of classical or \textit{ab initio} molecular dynamics simulations to treat the dynamics associated with large coverages and mixed species of adsorbates. While it would
be a challenge to perform \textit{ab initio} simulations of hundreds of adsorbates, it is be feasible to study the dynamics of 
small clusters or to  directly use more approximate semi-classical dynamical methods. The collective motion of multiple interacting adsorbates is expected to be significantly different from the independent-adsorbate dynamics studied here.

\ack
The authors thank Hartmut H{\"a}ffner, Dustin Hite, Jeremy Sage, Jonathon Sedlacek, John Chiaverini, and Jonathan DuBois for helpful discussions. 
This work was performed under the auspices of the U.S. Department of Energy by Lawrence 
Livermore National Laboratory under Contract DE-AC52-07NA27344. Part of this research was conducted using computational
resources and services at the Center for Computation and Visualization at Brown University.

\section*{References}

\bibliography{ref}{}

\end{document}

%\title{\textit{Ab Initio} Study of Electric Field Noise Heating in Ion Traps Caused by Electrode Surface Adsorbates}
\title[vdW-DF study of electric field noise heating in ion traps caused by adsorbates]{vdW-corrected density functional study of electric field noise heating in ion traps caused by electrode surface adsorbates: supplemental material}

\author{Keith G. Ray$^1$, Brenda M. Rubenstein$^{1,2}$\footnote{K. G. Ray and B. M. Rubenstein contributed equally}, Wenze Gu$^2$ and Vincenzo Lordi$^1$}

\address{$^1$ Quantum Simulations Group, Lawrence Livermore National Laboratory, Livermore, CA 94550, USA}
\address{$^2$ Department of Chemistry, Brown University, Providence, RI, 02912, USA}
\ead{ray30@llnl.gov}

%\date{\today}

% --------------------------------------------------------------------------------------

\section{Density functional theory calculations of interaction energies and adsorbate dipole moments}

In order to accurately model the interaction between the electrode surface and the adsorbates of interest, we utilize the Vienna Ab-Initio Software 
Package\cite{vasp1,vasp2,vasp3,vasp4} (VASP) with a plane-wave basis cutoff of 600 eV and pseudopotentials from the projector augmented
wave set\cite{paw2} (PAW).  The electrode is represented by an fcc gold (111) slab three atomic layers thick.  Because we employ periodic boundary conditions, thicker slabs were 
also tested. However, there was less than a 2\% difference in interaction energy between a six-layer slab and a three-layer slab.  Two different 
periodic cell in-plane sizes are used: a smaller cell (a 36-atom, three-layer slab) for atomic adsorbates, and a larger cell (a 144-atom three-layer slab)
for molecular adsorbates. Figure 1 depicts benzene adsorbed on the larger slab. A 1$\times$4$\times$4 gamma-centered $k$-point grid is used 
for the smaller cells and a 1$\times$2$\times$2 grid is used for the larger cells. The fcc Ag (111) surfaces were modeled just as the Au surfaces, but with a different lattice constant and pseudopotential. For the bcc Nb (110) surfaces we employ a three atom thick slab with 144 Nb atoms and a 1$\times$2$\times$2 gamma-centered k-point grid.
Periodic dipole interactions are corrected with the LDIPOL and DIPOL tags.\cite{PhysRevB.46.16067}  The exchange correlation functional 
we employ is the van der Waals density functional\cite{Dion:2004fk} with a ``consistently'' defined exchange functional (vdW-DF-cx).\cite{PhysRevB.89.035412}  
This functional is chosen to coherently treat the variety of adsorbates we consider, which include
both more strongly chemisorbed as well as physisorbed species, for which van der Waals interactions with the substrate dominate.  During 
ionic relaxations, forces are converged to better than 0.005 eV/\AA.  The fcc position was utilized for out-of-plane calculations on atomic adsorbates. For hydrogen on Au this yielded a binding energy of -2.47 eV, while the on-top and hcp positions yielded binding energies of -2.36 eV and -2.45 eV, respectively. CO only binds with the carbon closest to the Au surface and other orientations are unstable. Alternate adsorbate-substrate conformations explored for the hydrocarbon adsorbates demonstrated small energy differences, less than 10\% of the binding energy, due both to symmetries in the hydrocarbon molecules and the non-directional nature of van der Waals interactions.

\begin{figure}
\centering
\includegraphics[scale=1.1]{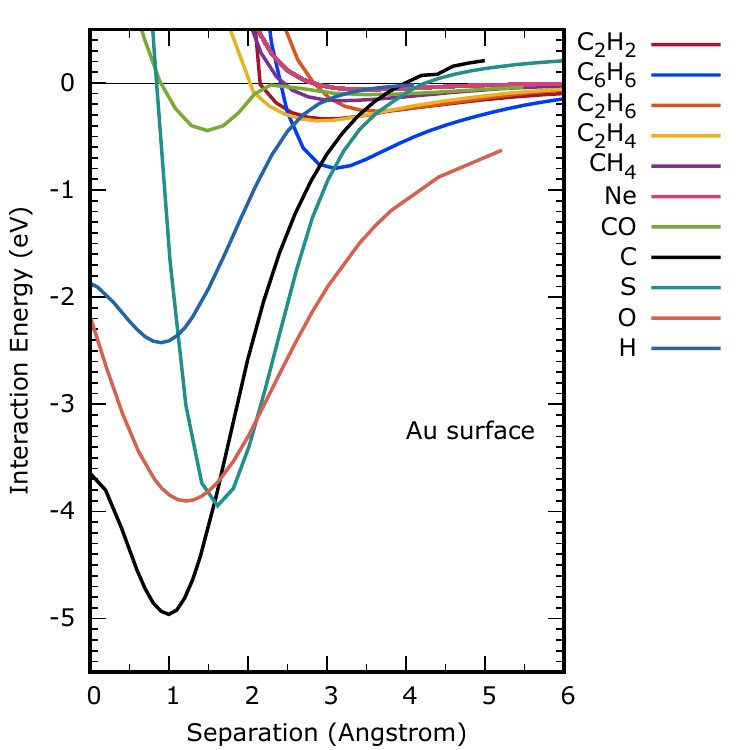}
\caption{Out-of-plane interaction potentials for adsorbates on the Au electrode surface.
}
\label{Aupot}
\end{figure}

\begin{figure}
\centering
\includegraphics[scale=1.1]{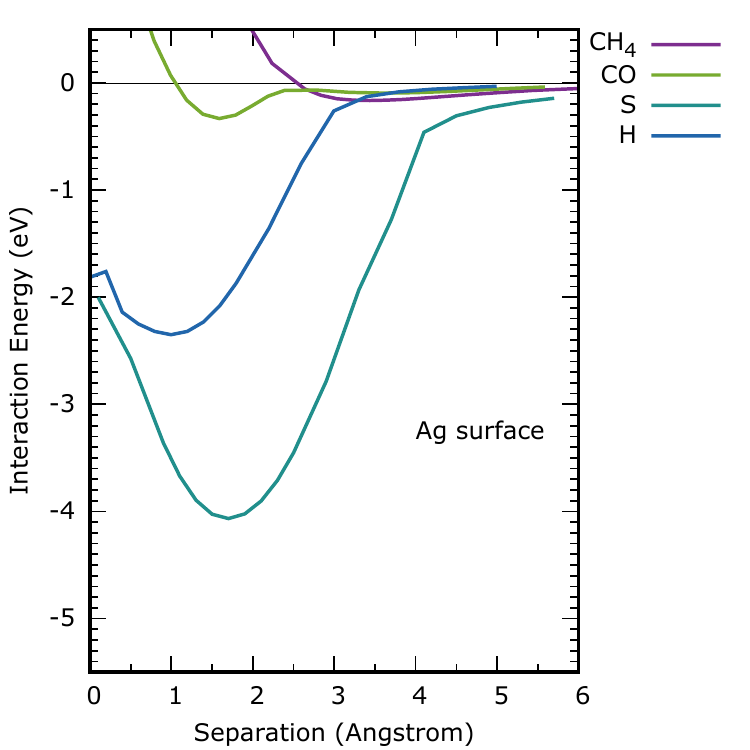}
\caption{Out-of-plane interaction potentials for adsorbates on the Ag electrode surface.
}
\label{Agpot}
\end{figure}

\begin{figure}
\centering
\includegraphics[scale=1.1]{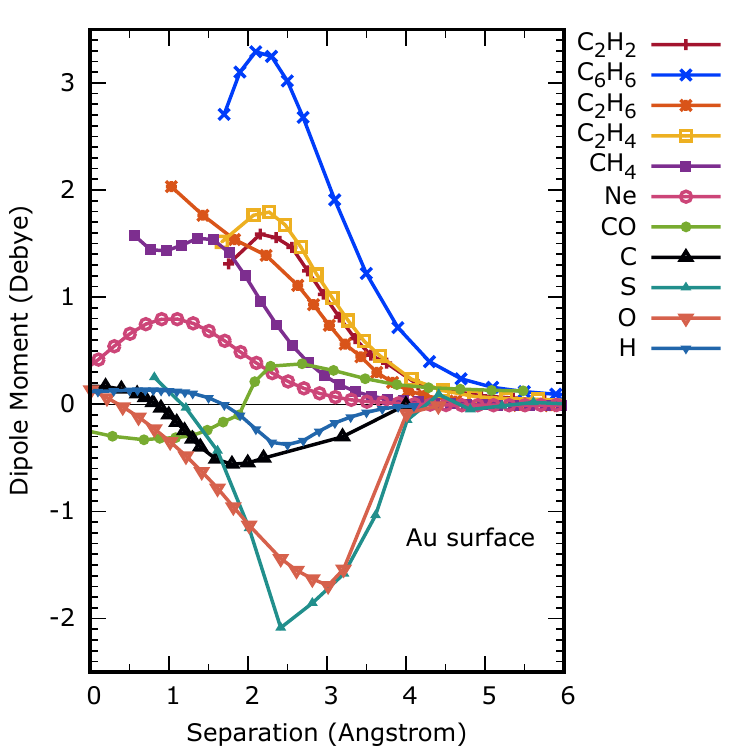}
\caption{Out-of-plane dipole moments for adsorbates on the Au electrode surface as a function of separation. Points are connected by lines as a guide to the eye.
}
\label{Audipole}
\end{figure}

\begin{figure}
\centering
\includegraphics[scale=1.1]{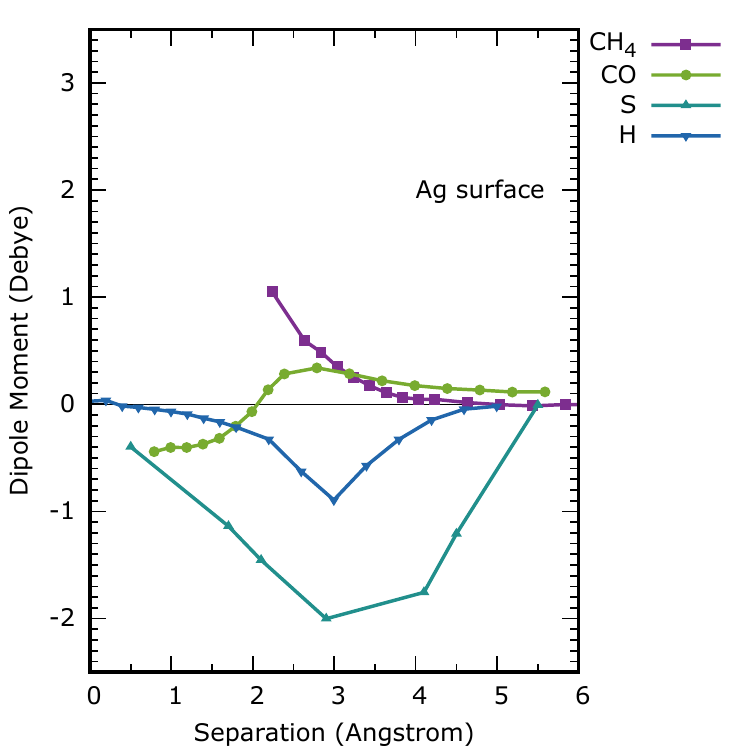}
\caption{Out-of-plane dipole moments for adsorbates on the Ag electrode surface as a function of separation. Points are connected by lines as a guide to the eye.
}
\label{Agdipole}
\end{figure}

\begin{figure}
\centering
\includegraphics[scale=1.1]{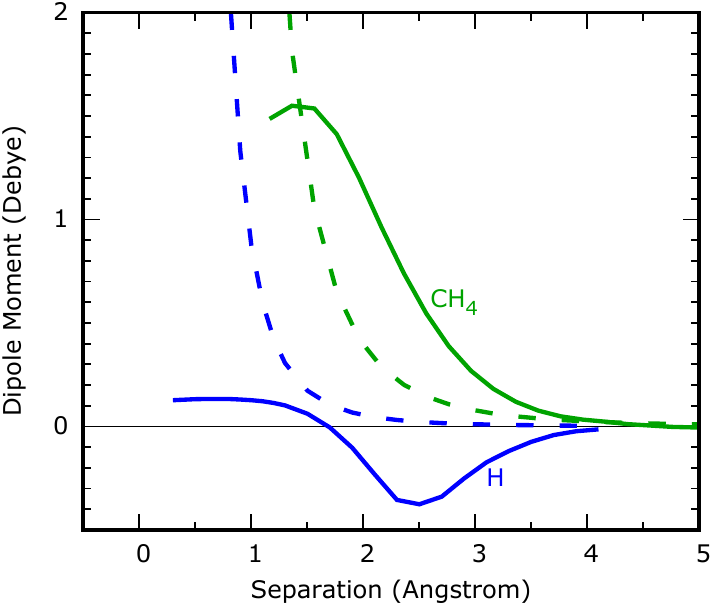}
\caption{Comparison of the previously proposed $z^{-4}$ functional form, dashed, for dipoles moments as a function of distance from the 
surface and DFT results, solid, for H--Au and CH$_4$--Au. 
}
\label{Figure4}
\end{figure}

Slab--adsorbate interaction energies are defined as
\begin{equation}
U_\textrm{int} = U_\textrm{slab+adsorbate} - U_\textrm{slab} - U_\textrm{adsorbate}.
\end{equation}
Interaction potentials for adsorbates on Au and Ag electrode surfaces are plotted in Figures \ref{Aupot} and \ref{Agpot}, respectively.  To calculate the dipole moment induced in the slab and adsorbate by their interaction, we integrate over the charge density output from VASP
using the standard formula,
\begin{equation}
\vec \mu = \int d^3 \vec r \, \delta \rho(\vec r) \vec r,
\end{equation}
where $\delta \rho (\vec r) $ is the difference between the sum of the isolated slab and adsorbate charge densities and the combined slab--adsorbate charge density. The adsorbate is far from the cell boundary and the change in the charge density has a vanishing monopole, so $\delta \rho(\vec r)$ is well-defined and 
independent of the origin. For these calculations, we use the HSE06\cite{:/content/aip/journal/jcp/125/22/10.1063/1.2404663} density functional to
correct for non-physical long distance charge transfer that can occur with local or semi-local exchange functionals.  Calculated dipole moments for adsorbates on Au and Ag electrode surfaces are plotted in Figures \ref{Audipole} and \ref{Agdipole}, respectively.  Previous works have used a different form for $\mu_{z}(z)$,\cite{safavi2011microscopic}
\begin{equation}
\mu_{z}(z) = 0.47 e a_{0}^{1/2} \alpha^{3/2} z^{-4}, 
\end{equation}
where $\alpha$ is the adsorbate polarizability. 
As illustrated in Figure \ref{Figure4} for H and CH$_4$, this previous form can be substantially different from the forms we obtain by direct integration of the charge density obtained through DFT calculations.


\section{Calculation of average dipole moments}

In order to determine the state-averaged dipole moments that feed into our dipole-dipole correlation functions, we fit DFT-based 
results for a given adsorbate's induced dipole moment as a function of height with a smooth curve, $\mu(z)$. We then compute the average dipole 
moment in state $i$, $\mu_{z,i}$, as

\begin{equation}
\mu_{z,i} = \langle i|\mu_{z}(z)|i \rangle 
\end{equation}
by numerically integrating the induced dipole moment over wave function $i$. We obtain the energies and wave functions of the adsorbate vibrational states by numerically solving the 1D Schr\"{o}dinger equation.

\section{Simulation of dipole-dipole correlation functions and spectra}

According to the patch potential model, heating rates are directly proportional to the dipole-dipole fluctuation spectrum
\begin{equation}
S_{\mu}(\omega) = \int_{-\infty}^{\infty} d\tau [\langle \mu_{z}(\tau) \mu_{z}(0) \rangle - \langle \mu_{z}(0) \rangle^{2} ]e^{i \omega \tau}, 
\end{equation}
which in turn is the Fourier Transform of the dipole-dipole correlation function
\begin{equation}
C_{\mu}(\tau) = \langle \mu_{z}(\tau) \mu_{z}(0) \rangle. 
\end{equation}
The total dipole moment, $\mu_{z}$, may be written as the sum of contributions from each vibrational state
\begin{equation}
\mu_{z} = \sum_{i} \mu_{z,i} \rho_{i}, 
\label{DipoleMomentEquation}
\end{equation}
where $\rho_{i}$ denotes the population of vibrational state $i$. At equilibrium, $\{\rho_{i}\}$ are given by Boltzmann factors. The dipole-dipole correlation function may therefore be re-expressed as
\begin{equation}
C_{\mu}(\tau) = \sum_{ij} \mu_{z,i}  \mu_{z,j}  \langle \rho_{i} (\tau)\rho_{j} (0)\rangle.
\label{CorrelationFunctionEquation}
\end{equation}

We compute this correlation function by solving the master equation for the time-dependent populations of the vibrational states
\begin{equation}
\frac{d}{dt} \langle \rho_{i} \rangle = \sum_{j} M_{ij} \langle \rho_{j} \rangle,  
\end{equation}
where $M$ is the matrix of transition rates in which 
$M_{ii} = - \sum_{j \neq i} \Gamma_{i \rightarrow j}$ and $M_{ij} = \Gamma_{i \to j}$ for $i \neq j$.\cite{safavi2011microscopic} We solve this master equation via kinetic Monte Carlo simulations.\cite{Bortz_1975,Gillespie_1976} In specific, we distribute our walkers evenly among the different vibrational states, and then, as is typical in kinetic Monte Carlo simulations, allow these populations to fluctuate according to the Monte Carlo dynamics until equilibrium is achieved.

We do this by first calculating the transition probabilities 
among states, $\omega_{i \rightarrow j}$,
\begin{equation}
\omega_{i \rightarrow j} = M_{ij} \rho_{i}
\end{equation}
and the total transition probability for leaving a given state $i_{0}$
\begin{equation}
W_{i_{0}} = \sum_{i_{0} \neq j} \omega_{i_{0} \rightarrow j}. 
\end{equation}
Based upon these quantities, we then determine the time interval until the next walker jumps from one vibrational state to the next, 
$t_{i_{0} \rightarrow j}$, using
\begin{equation}
t_{i_{0} \rightarrow i_{1}} = \frac{-\ln u_{0}}{W_{i_{0}}}, 
\end{equation}
where $u_{0}$ represents a random number drawn from a uniform distribution. Once we know the
time until the next jump, we then determine to which new state a walker in state $i_{0}$ will jump. The probability of jumping to state
$i_{1}$ from state $i_{0}$, $p_{i_{0} \rightarrow i_{1}}$, given that a jump will occur is given by 
\begin{equation}
p_{i_{0} \rightarrow i_{1}} = \frac{ \omega_{i_{0} \rightarrow i_{1}}}{W_{i_{0}}}. 
\end{equation}
We sample this distribution by drawing a second random number, $u_{1}$, from a uniform distribution and choosing the smallest state, $i_{1}$, such that 
\begin{equation}
\sum_{j=1}^{i_{1}} p_{i_{0}\rightarrow j} > u_{1}.
\end{equation}
Once the final destination of the walker is known, we update the time to $t_{1} = t_{0} + t_{i_{0} \rightarrow i_{1}}$ and move the walker to 
state $i_{1}$. We iterate this process until equilibrium is achieved, as evidenced by the populations in each state settling to their Boltzmann 
values. During the course of the kinetic Monte Carlo simulations, we also compute the average dipole moment (Equation \ref{DipoleMomentEquation}) 
after each jump and use this information to later calculate the dipole-dipole correlation functions (Equation \ref{CorrelationFunctionEquation}).

After obtaining the dipole-dipole correlation functions in this manner, the related spectra are obtained via 
a numerical Fourier Transform of the correlation functions. We perform this transformation using the Fast Fourier Transform (FFT)
algorithm.\cite{press1989fast} Occasionally, the correlation functions to be transformed are too noisy to yield smooth spectra. In such cases, the tails of the correlation functions, which are typically the noisiest portion, are first cut off and then the remaining portion of the correlation functions is filtered using SciPy's Savitzky-Golay filter.

The electric field spectra, $S_{E}(\omega)$, and heating rate, $\dot{\hat{n}}$, may then be obtained by the expressions described in the main text. The heating rates calculated in this manner are presented in Figure 4  of the main text. 
\begin{table*}
\begin{tabular}{l*{5}{c}}
%\begin{ruledtabular}
%\begin{tabular}{r.{2.5}.{2.5}.{2.5}.{2.5}.{2.5}.{2.5}}
\hline
& \multicolumn{5}{ c }{Residence time (s) at temperature (K)}\\
Adsorbate on Au & 100 K & 200 K & 295 K & 400 K & 500 K \\
\hline
C$_2$H$_2$       &  5.3 $\times$ 10$^4$  &   1.8 $\times$ 10$^{-4}$ &   3.4 $\times$ 10$^{-7}$  &   1.1 $\times$ 10$^{-8}$&   1.5 $\times$ 10$^{-9}$  \\
C$_6$H$_6$       &  7.0 $\times$ 10$^{27}$  &   5.8 $\times$ 10$^{7}$ &   2.0 $\times$ 10$^{1}$  &   5.3 $\times$ 10$^{-3}$&   5.2 $\times$ 10$^{-5}$  \\
CH$_4$       &  1.1 $\times$ 10$^{-4}$  &   7.5 $\times$ 10$^{-9}$ &   3.4 $\times$ 10$^{-10}$  &   6.2 $\times$ 10$^{-11}$&   2.4 $\times$ 10$^{-11}$  \\
S     &  1.5 $\times$ 10$^{186}$  &   5.5 $\times$ 10$^{86}$ &   5.3 $\times$ 10$^{54}$  &   1.1 $\times$ 10$^{37}$&   1.2 $\times$ 10$^{27}$  \\
CO     &  4.5 $\times$ 10$^{9}$  &   2.9 $\times$ 10$^{-2}$ &   7.2 $\times$ 10$^{-6}$  &   7.4 $\times$ 10$^{-8}$&   5.6 $\times$ 10$^{-9}$  \\
\hline
\end{tabular}
%\end{ruledtabular}
\caption{Selected adsorbate residence times on the electrode surface.}
\label{Table9}
\end{table*}

\section{Desorption rates}

In order to assess the relative likelihood of the presence of different adsorbates on the electrode surface, we have calculated adsorbate residence times at different temperatures for representative species.  To accomplish this, we calculate the desorption rate according to the Arrhenius or Polanyi-Wigner rate equation for a first order reaction,\cite{redhead1962thermal} $k_\textrm{des} = A\times e^{E_\textrm{binding}/kT} $, where $A$ is the vibrational frequency of the adsorbate out-of-plane vibrational mode, $E_\textrm{binding}$ is the binding energy of the adsorbate (defined to be negative), and kT is the Boltzmann constant multiplied by the temperature.  The average residence time is given by $T_\textrm{res} = \frac{1}{k_\textrm{des}}$ and the results are presented in Table \ref{Table9}.  The qualitative order of magnitudes of the residence times span an enormous range and therefore this calculation should be a meaningful indication of the relative residence time by adsorbate, even if it is not quantitatively predictive.

% --------------------------------------------------------------------------------------

% --------------------------------------------------------------------------------------
\ack

This work was performed under the auspices of the U.S. Department of Energy by Lawrence 
Livermore National Laboratory under Contract DE-AC52-07NA27344. Part of this research was conducted using computational
resources and services at the Center for Computation and Visualization, Brown University.

\section*{References}

\bibliography{ref}{}